\documentclass{iucrjournals}
\usepackage{lineno}
\nolinenumbers

\usepackage{xcolor}
\usepackage{siunitx}
\title{Understanding ultrafast x-ray "echoes" diffracted from single crystals}

\author[1,*,+]{Angel Rodriguez-Fernandez}
\author[2]{Dmitry Karpov}
\author[2]{Steven Leake}
\author[3]{Dina Carbone}
\author[4]{Ana Diaz}

\affil[1]{European XFEL Facility GmbH, Holzkoppel 4,Schenefeld, DE-22869}
\affil[2]{ESRF, Grenoble, F-38000}
\affil[3]{MAX IV Laboratory, Lund University, Lund, SE-22100}
\affil[4]{PSI Center for Photon Science, Paul Scherrer Institute, Forschungsstrasse 111, Villigen PSI, CH-5232}

\affil[*]{Corresponding author: angel.rodriguezfernandez@dimamond.ac.uk}
\affil[+]{New affiliation: Diamond Light Source, Harwell Science and Innovation Campus, Fermi Avenue, Didcot, UK OX11 0DE}

\begin{document} 
\maketitle 

\begin{synopsis}
In the diffraction peak of highly perfect crystals, it is possible to observe a fine structure of maxima and minima, generated from the volume from which the x-rays are diffracted, known as the Borrmann fan.
These multiple beams, also known as echoes, share the same monochromatic properties, are parallel to each other, and travel with a short delay of a few femtoseconds between each other. 
The echoes structure inside the Borrmann fan is generated by the interference of the multiple diffracted beams along the crystal thickness at the exit crystal surface. 
Here, we characterize the fine structure of a diffraction peak generated by a Si \SI{100}{\micro\meter} thick crystal with a resolution of about about \SI{100}{\nano\meter}.
\end{synopsis}

\begin{abstract}
Multiple x-ray beams generated by interference processes in perfect crystals were imaged with a resolution of about \SI{100}{\nano\meter} using tele-ptychography in the diffraction direction.
%Generation of ultrafast multiple x-ray beams have been observed in a Si crystal in the diffraction direction with high resolution. 
%The experiment was perform using the high coherent flux provided by 4th generation synchrotron sources to image the wave-field. 
These multiple wave-fields, also known as x-ray diffraction echoes, are related to the process known as the Pendell\"{o}sung effect and are described by dynamical diffraction theory. 
The echoes are produced by the constructive interference of diffracted x-rays at the exit surface of the crystal sample.
In the imaged diffraction peak, we observed 10 echoes maxima with a total signal length of \SI{78}{\micro\meter}.
%The echoes are spatially displaced by less than \SI{78}{\micro\meter}, 
Which translates into a total temporal delay in the signal of less than \SI{108}{\femto\second}.% for the full signal.
%following the work by Shvydko and Lindberg \cite{Shvy12}. 
This makes the echoes of high importance for x-ray optics at x-ray Free Electron Laser sources, as the effect could be used for future ultrafast x-ray beam splitters.  
In addition to this application, echoes can be exploited to follow ultrafast processes in single crystal micro-structures such as melting or strain propagation.
%This is because the echoes show a modulation with the lattice deformation along the depth in the time that the x-rays needs to propagate and diffract along the crystal at the speed of light.
\end{abstract}


%XFELs are great instruments for time-resolved structural experiments thanks to the small duration pulses it generates with many photons
The emergence in the last decade of hard x-rays free electron laser facilities (XFELs) \cite{LCLS10,EUXFEL20} is enabling ultrafast science at a new frontier.
The x-ray pulses generated at these facilities have a high transversal coherence and a temporal length below \SI{200}{\femto\second}.
These two properties are perfect for imaging ultrafast processes in matter for studies of liquid, powder or solid samples.
%pump-probe
Examples of these studies are imaging of cavitation bubbles generated with a femtosecond laser in a pump-probe setup \cite{salditt21}, understanding the molecular dynamics via x-ray photon correlations spectroscopy \cite{Reiser22,Dallari21}, reconstruction of surface melted nano-particles using coherence diffraction imaging \cite{Robinson20} or study of ultrafast melting and shock wave propagation in metals or semiconductors \cite{Zalden19,Marais23,ANTONOWICZ2024,Irvine25}.
%For very high resolution experiments, it is difficult to generate pulses separated by short intervals of a few tens of femtoseconds in such a way that the timing of the pulses is precisely known 
These new facilities promise a better understanding of ultrafast x-ray processes for x-ray optics that could not be measured at synchrotron sources, because the temporal length of the x-ray pulses is on the order of tens or hundreds of picoseconds.
In order to exploit the great potential of XFELs for imaging of ultrafast processes in the femtosecond time scale, it would be optimal to create trains of pulses with separation in the femtosecond range. 
A way to achieve this is by using splitting and delay lines (SDL), that provide delay times of few femtoseconds \cite{Roseker2009,Roseker2011,roseker2018,Lu2018} between x-ray pulses. 
However, for the investigation of faster processes, even shorter pulse lengths would be needed.
Additionally, the diffraction from an asymmetric or strained crystal has been proposed as a way to shorten the x-ray pulse length \cite{CHAPMAN2002,Christov2006,Hrdy2013}.
This motivates the exploration of ultrafast x-ray effects in perfect and moderately strained crystals, for suitable time-resolved beam splitters and experiments at sub-femtosecond x-ray facilities.
%Not only for a more stable pulse to pulse performance, but also for an on-demand pulse temporal length, with shorted time delays between pulses as in the case of the splitting and delay lines \cite{SDL} or even shorter pulses by using asymmetric cut crystals \cite{asymetric_crystal}.

%But to improve it is also need to evolve optics and understanding on basic ultrafast diffraction problesms.

%more talk about the XFELs
%Introduce the optical problem
%Diffraction ultrafast effect
%The echoes produced in the process of x-ray diffraction by perfect crystals have very short time intervals of a few tens of femtoseconds, as determined by calculations. Therefore it is interesting to explore this phenomenon and characterize it for its use as a probe for XFEL experiments.

%link these two things
During the 1960s there was a significant increase in our understanding of dynamical diffraction, including the exploration of the Borrmann fan \cite{Borrmann49,Shull68,Shull86,Authi01,Auth12}.
Borrmann observed that x-rays diffracted by crystals in Laue transmission geometry extended more than the size of the incident beam.
He observed that the exit fingerprint extended along the crystal exit surface , for both the forward and the diffracted beams. 
This structured intensity signal, known as the Borrmann fan, is governed by the Pendell\"{o}sung effect as described in dynamical diffraction theory. 
In Laue geometry, under diffraction conditions, the Pendell\"{o}sung length is the thickness of the crystal needed to change beam propagation from the forward to the diffraction directions, similar to the concept of extinction in Bragg geometry.
Greater understanding of the Borrmann effect came from neutron diffraction and interferometry based in arrays of silicon crystals \cite{Shull68,COW75,Shull86,Arthur85a}.
Neutron dynamical diffraction based interferometers have been used recently to resolve with higher resolution the forces of the universe \cite{Heacock22}.
Back to x-rays, simulations based on dynamical diffraction theory showed that the generated beams are not continuous, but consist of a series of spatially displaced and temporally delayed beams.
These x-ray beams travel parallel to each other \cite{Bushu08,Shvy12}, similar to sound echoes, but in the electromagnetic range.
It is only now, with modern synchrotron sources that we have obtained enough spatial resolution to observe the Borrmann fan, as presented in \cite{Carlsen22}.
%with high resolution.
Finally, the new ultrafast sources, such as European XFEL, allow the study of the temporal response in the order of the femtosecond.

%describe Borrmann fan
To better comprehend the process in which the dynamical diffraction echoes are generated, we have to understand the Borrmann fan in detail.
In the diffraction condition the crystal acts as a beam splitter dispersing x-ray momenta inside the reflection width to trajectories within the Borrmann fan.
Such a plane wave momentum state creates 2 wave-field solutions inside the crystal traveling one in the forward and one in the diffraction directions.
These two wavefronts propagate along the crystal towards the back surface.
%, where each combines with other overlapping components. 
Inside this fan, the x-ray beam is not localized and acts like a superposition of quantum states of probability for each photon.
The x-ray wave-fields are coherent inside the crystal within this fan.
%The echoes are monochromatic as during the process only one reflection is being excited.
The interference between these components creates the exit beam modulation structure, Pendell\"{o}sung, both in the diffraction,$k_H$, and forward, $k_0$, directions.
In this way, we can say that the crystal acts as a beam splitter \cite{Shull86}.
%This effect is known also as spatial Pendell$"o$sung oscillations
%, due to it all echoes share the same energy bandwidth.
The echoes produced in the x-ray diffraction process have very short delays between each other, on the order of a few tens of femtoseconds, depending on the sample thickness and Pendell\"{o}sung length \cite{Shull68}.%,  a few tens of femtoseconds. 
%The delays between pulses are in the order of the $fs$.
The overall dynamical diffraction echoes present a temporal structure that lasts less than \SI{108}{\femto\second} in the case of a thin Si crystal of \SI{100}{\micro\meter}. 
This temporal structure can only be resolved with the short pulses provided by attosecond or femtosecond x-ray sources.
%As an ultrafast phenomena exploring the echoes is interesting for its future at XFEL experiments to generate on demand pulse structures in splitting and delay lines schemes.

%present the publication
In the past, we achieved direct measurement of echoes in the forward direction\cite{ARF18,ARF20,ARF21}.
The spacing of the echoes could be measured with high spatial resolution.
From these measurements the time delays could be inferred to match the expected values from the calculations.
However, in the forward direction echoes decay in intensity very rapidly across the Borrmann fan, which is not ideal if one wants to use them as probes or splitting generators at XFELs.
In particular, we have shown how a plastic deformation on the surface of a silicon crystal affects the position and intensity of the echoes \cite{ARF21}.
More recently, it has been presented how the strain in a crystal can modulate the signal of the dynamical diffraction fringes in Bragg geometry \cite{Krzywinski22}.
This modulation together with the spatio-temporal coupling of the echoes shows that the echoes can be used to probe ultrafast deformations happening along the crystal depth, as each echo relates to the coherence interference of beams probing a defined area in the Borrmann fan.
In a recent work, we have presented the effect on the echoes of a femtosecond laser excitation on the front surface of a Si single crystal in an experiment performed in a pump-probe geometry at the European XFEL \cite{ARF25}.
%\textcolor{blue}{Talk about delay depth and strain with citation to PRL \cite{ARF21}}. 
%talk about absorption

In this work, we have extended our study to the diffraction direction signals in Laue geometry and imaged the structure of the echoes produced by a Si wafer with \SI{100}{\micro\meter} thickness with a resolution of about \SI{100}{\nano\meter}.
We have measured static echoes in the diffraction direction in a synchrotron in preparation for future time-resolved experiments at XFELs.
Each of the echoes measured in the diffraction direction shows similar intensity with respect to the others, as expected from simulations.
The wavefront was imaged at the ID01 beam line of the ESRF EBS facility using tele-ptychography \cite{Tsai26}.
With this variant of ptychography \cite{Pfeiffer18} it is possible to image complex wavefronts after propagation through a sample in cases where the assumption of a factorization of the illumination and the sample transmissivity would not work, which is the case for dynamical diffraction.
 Additionally, tele-ptychography allows one to perform very local measurements by focusing the incoming beam on a spot of the sample and in this way probe the local lattice distortion as performed in \cite{ARF21}. 
 Although the wavefront measurement is performed downstream of the sample, where the beam is divergent, back propagation allows the retrieval of the echoes at the focal plane, where it does not overlap with any other signal. 
 Therefore, tele-ptychography is a perfect tool for a clean measurement of dynamical diffraction fringes produced by thin crystals.

%comutation of the sample and the wavefront
%That are systems that do not work in the kinematical diffraction regime, and in which dynamical diffraction theory must be applied, such in the case of the echoes.
%An example of these complex wavefronts are the echoes, which can only be described using dynamical diffraction theory.
%Obtaining an spatial resolution of \textcolor{blue}{\SI{50}{nm}}.
The separation between the echoes in this particular work ranges from about \SI{2}{\micro\meter} to \SI{78}{\micro\meter}, which corresponds to a temporal delays of 2 fs to 108 fs.
%As it is presented in this work, the maximum displacement between the different echoes, $\Delta$x, measured is around \SI{78}{\micro\meter}.
%This would correspond to a maximum temporal delay, $\Delta$t, of \SI{108}{\femto\second} between the first and the last echo.
Today, short delays between echoes are difficult to probe at XFELs, because the temporal resolution in the actual facilities is in the order of 10 to \SI{100}{\femto\second}.
In the last years, the scientific community has shown interest in following electron dynamics in atoms, molecules and nanoscopic systems \cite{Krausz2009}.
As these processes happen in the attosecond time scale the accelerator scientist community has been driven to the design of new XFELs concepts for the production of shorter x-ray pulses \cite{SALDIN2004,Zholents2005,Prat2015,duris2020}.
In this sub-femtosecond facilities the effects of the echoes would be easier to use in the understanding of ultrafast processes.
%\textcolot{blue}{in thin crystals or used to probe with really short delays samples by many beams in an beam splitting scheme.}

%Talk about the effect
\section{Experiment and data collection}
%\subsection{Experimental setup}

\begin{figure}[htbp]
\centering
\fbox{\includegraphics[scale = 0.16]{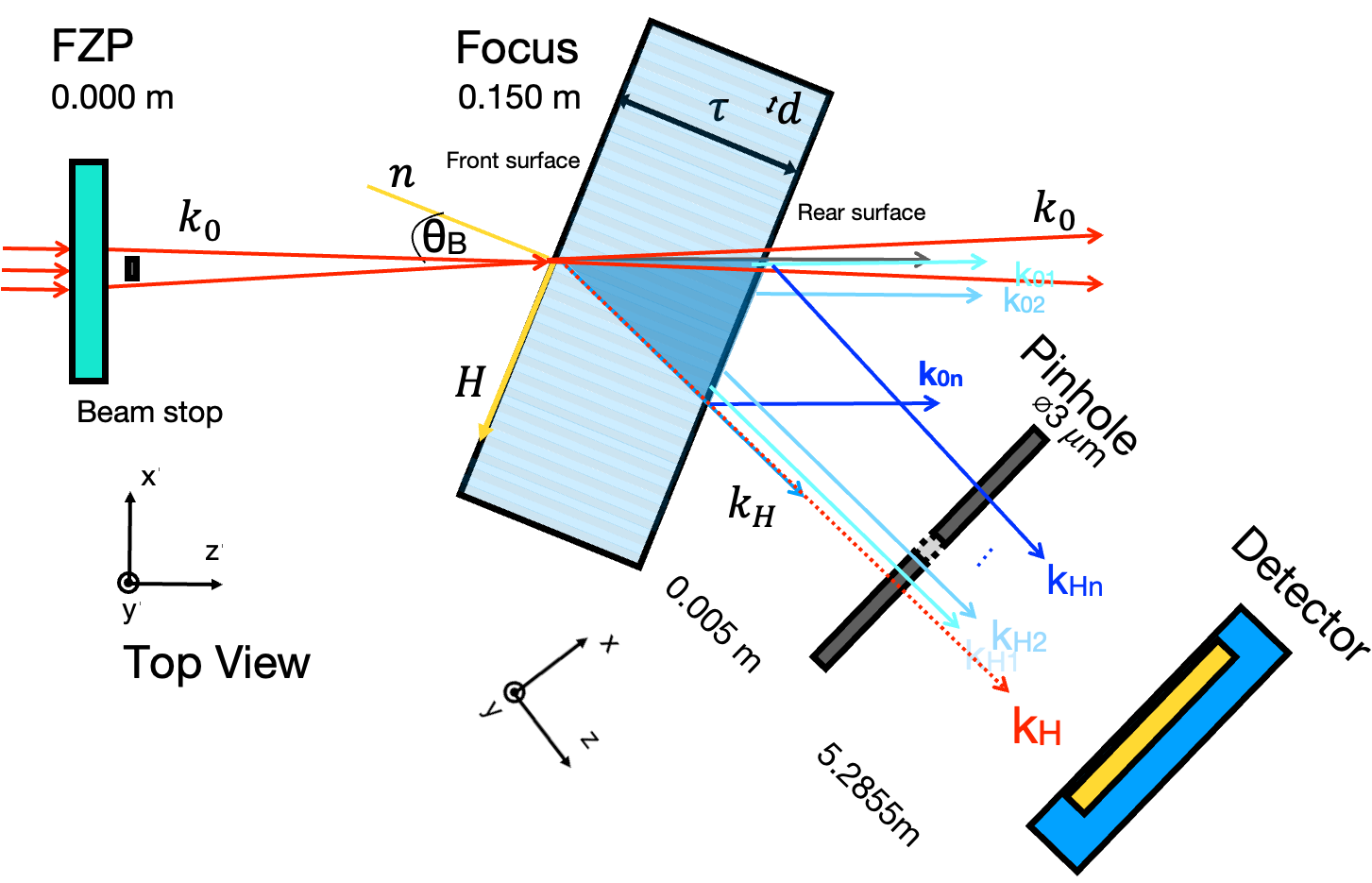}}
\caption{Sketch of the experimental geometry. 
The incoming x-ray beam was focused with a Fresnel zone plate (FZP) to \SI{73}{\nano\meter} onto the center of rotation of the goniometer. 
The Si wafer was positioned at the focus and set to the (220) Laue diffraction condition in the horizontal plane. 
A pinhole with diameter \SI{3}{\micro\meter} was located at \SI{5}{\milli\meter} from the rear surface of the crystal. 
The distance between sample and detector was of \SI{5.285}{\meter}. 
%maybe comment SU of the pinhole?
}
\label{fig:Geometry}
\end{figure}

The experiment was performed at the nano-diffraction beamline ID01 \cite{Leake19}, at the ESRF-ESB facility \cite{ESRFEBS16}.
The sample under study was a \SI{100}{\micro\meter} thick Si wafer with its surface oriented perpendicular to the $(001)$ plane family.
%On a small area of the sample micro-indentations were performed a diamond Berkovich \SI{1}{\micro\meter} tip with loads between \SI{10} and \SI{75}{\milli\newton}.
%In between indents was left a distance of \SI{20}{\micro\meter}.
%This indentations were performed in-situ using a SEM nanoindenter, Zeiss Leo Ultra 55 FEG. 
%The indentations were performed at Chalmers university of Technology.
%In this experimental study 
The reflection under study was the (220) in horizontal geometry, which is presented in Fig. \ref{fig:Geometry}.
This is a co-planar Laue symmetric reflection with a Bragg angle, $\theta_B$, of \SI{22.75}{\degree} at \SI{8.346}{\kilo\electronvolt},
and an extinction length of \SI{23}{\micro\meter}, which sets a lower limit in thickness for which dynamical diffraction effects can be observed.

%The sketch of the experiment is presented in Fig \ref{fig:Geometry}.
The energy was selected by a Si (111) double-bounce vertical monochromator, at this energy the coherent photon flux is $10^{10}\,$photons/s, where the coherent part of the beam was selected in the transverse direction by setting the slits before the focusing optics to \SI{500}{\micro\meter} x \SI{140}{\micro\meter} in the vertical and horizontal directions, respectively.
%\textcolor{blue}{size slits}.
A \SI{300}{\micro\meter} diameter Fresnel Zone Plate (FZP), with a focal distance of \SI{121.2}{\milli\meter}, was used to focus the wavefront to the center of rotation of the diffractometer.
After the FZP the central part of the x-ray was blocked by a \SI{60}{\micro\meter} beam-stop.
These parameters allowed for a beam size at the focus of \SI{73}{\nano\meter}, %, translates in a divergence of \SI{3.9}{\milli\radian}.
with a depth of focus of about \SI{97}{\micro\meter}, nearly equal to the thickness of the sample under study.
Because the crystal was placed at the beam focus, we can assume that both surfaces are inside the depth of focus, providing the best spatial resolution for studying dynamical diffraction fringes generated at the exit surface of the crystal.
In addition, the focused beam divergence excites a multitude of states within the Borrmann fan, increasing the signal compared to plane-wave illumination.
%Since the crystal was placed in the focus of the beam, in that way we can assume that the x-ray beam is at focus for the two surfaces of the crystal.
%The central depth of the Si crystal was located in the center of the depth of focus, in that way we can assume that the x-ray beam is at focus for the two surfaces of the crystal. 
%This beam size at the focus provides the best resolution for the description of the position and helps to discriminate the position of the dynamical diffraction fringes generated by the sample.
%Due to the focusing the divergence of the incident beam excites the Borrmann fan,i.e. it allows more photons to fulfil the diffraction condition, what translates in more signal in the fringes that in the case of the plane wave.

An Eiger 2M detector, developed and manufactured in a collaboration between ESRF and the Paul Scherrer Institute, was located at the $2 \theta_B$ angle in the horizontal geometry at a distance of \SI{5.285}{\meter}.
In between the sample and the detector we placed a \SI{3}{\micro\meter} pinhole fabricated on a \SI{50}{\micro\meter} thick tungsten foil. The pinhole was mounted on a NPXY50-286 Nanopositioning Piezo Stage from nPoint for scanning.
%As sketched in Fig.\ref{fig:false-color}
The distance between the pinhole and the sample was \SI{5}{\milli\meter}.
%, while between pinhole and the detector was of \textcolor{blue}{\SI{5.545}{\meter}}.

%\subsection{Data Analysis}

%\begin{figure*}[htbp]
\begin{figure}[htbp]
\centering
\includegraphics[scale = 0.4]{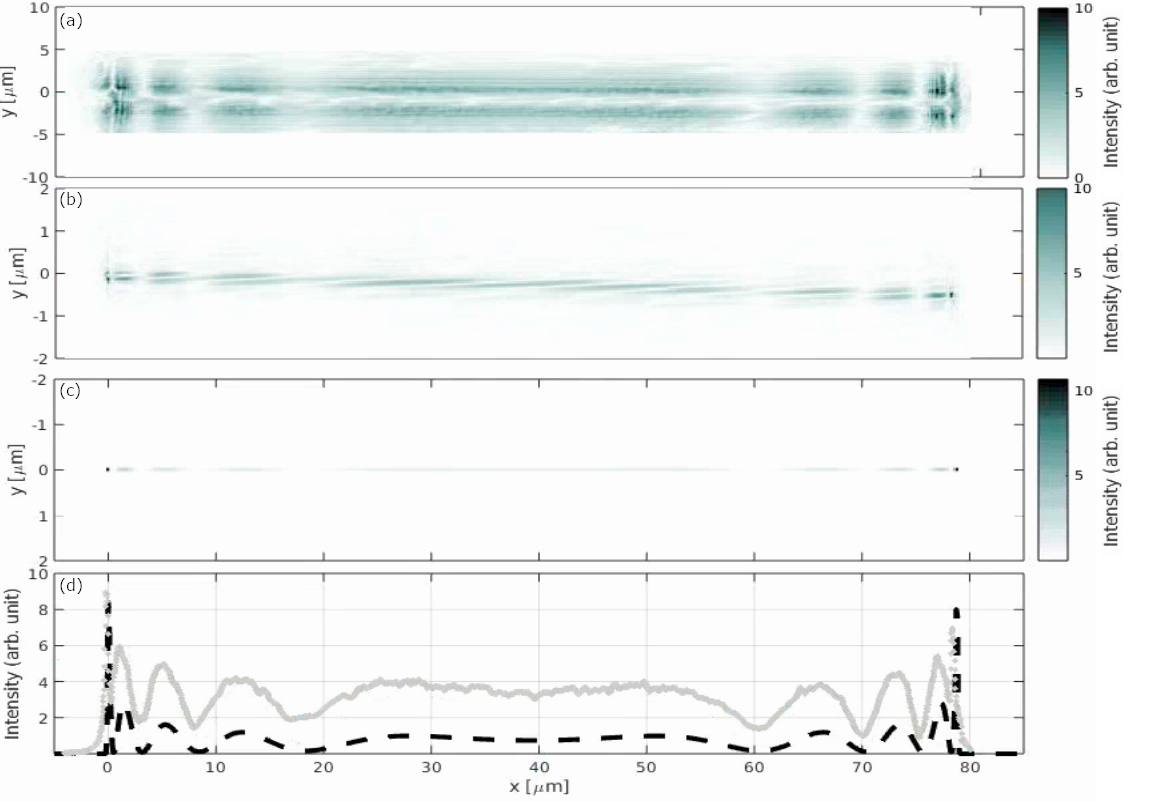}
\caption{
(a) Intensity of the reconstructed wavefronts at the pinhole plane for the (220) reflection of a \SI{100}{\micro\meter} Si thick crystal at \SI{8.346}{\kilo\electronvolt}.
(b) propagation of the reconstructed wavefront back to the rear surface of the crystal.
(c) Simulated dynamical diffraction wavefront of the Laue symmetric $(220)$ reflection for a \SI{100}{\micro\meter} thick crystal at \SI{8.346}{\kilo\electronvolt}
(d) projection along the y direction of the measured diffraction intensity in (b) (black stars) and simulated dynamical diffraction signal (dash line).
}
\label{fig:reconstruction_all}
%\end{figure*}
\end{figure}

%Collection of the data
The intensity of the diffracted beam, for the Si $(220)$ Laue reflection, was collected by performing a series of tele-ptychographic scans \cite{Tsai26,verezhak2020,ARF21}.
The tele-ptychography data collection was performed by scanning the pinhole in the x-y plane (normal to the propagation direction) following a spiral trajectory with a step size of about \SI{570}{\nano\meter} in the vertical and horizontal directions.
To reduce scanning time, the spiral was limited to \SI{20}{\micro\meter} in the vertical direction, keeping the full range of the piezo stage, \SI{50}{\micro\meter}, in the horizontal direction.
In order to collect the full signal, due to the short range of our scanner (\SI{50}{\micro\meter}), two adjacent areas were scanned trying to keep and overlap in between the two scanned areas of \SI{10}{\micro\meter}, as presented in the supplementary material (SM) Fig. S1.
%, we denote the second area scanned as (').
A coarse motor was used to move between the two scan areas with a step of \SI{40}{\micro\meter} following the direction of the diffracted signal along the x-axis. 
%In this way, we ensured an overlapping area of  width between the two scanned areas.
%The step of \SI{55}{\micro\meter} was chosen to leave enough empty areas around the diffracted signal, so the propagation to the focus could work without aberrations.
%Unfortunately this step was to big to record all the signal, and the newport \textcolor{blue}{model} did not allow for a high reproducibility.

%During the experiment the data was reconstructed using the PyNX software package \cite{FN_pyNX20}, with a combination of 10 iterations of the difference map (DM) algorithm \cite{Thibault09} without updating the reconstruction of the probe and 1200 iterations of DM allowing probe update, followed by 1000 iterations of a maximum likelihood (ML) refinement \cite{Thibault12} with update for both wavefront and probe.
%Finally we performed a refinement with 800 ML extra iterations in which the position of the pinhole was allowed to vary.
%With this the diffracted wavefront at the pinhole plane was reconstructed.
Both scans were used simultaneously in a joint ptychographic reconstruction sharing the object, while allowing one reconstructed probe for each of the scans, as described in Guizar-Sicairos et al. \cite{Guizar14}.
The ptychographic reconstructions were performed using the PtychoShelves package developed in Matlab by the Coherent X-ray Scattering group at the Paul Scherrer Institute in Villigen Switzerland \cite{Wakonig20}.
%This routine applies a common reconstruction of the wavefront in the 2 combined scanned areas, while the pinhole refinement is independent for the two scans \cite{Guizar14}.
The algorithm started with 10 difference map (DM) iterations in which the probe of the reconstruction (the pinhole) was fixed from an initial guess \cite{Thibault09}, presented in Fig. S2 of the SM \cite{SM20}.
This initial guess probe was obtained from a tele-ptychographic reconstruction performed in the forward  direction for the same diffraction condition, see Fig.~S3 of the (SM) \cite{SM20}, as previously described in our work \cite{ARF21}.
After this initial 10 iterations, both the object and the probe were allowed to vary for 90 more iterations of the DM algorithm.
Finally, we added 500 iterations of a maximum likelihood (ML) refinement \cite{Thibault12} with update of both, wavefront and pinhole, as presented in Fig. S4 from the SM \cite{SM20}.
During the reconstructions, 3 different modes were used for the probe (pinhole) as presented in Fig. S5 of the SM\cite{SM20}

%In the reconstruction we use the pinhole obtained with a test specimen as initial guess, see Fig.~S4 of the SM \cite{SM20}.

\section{\label{sec:level3} Results and Discussion\protect}
%\begin{figure}[htbp]
%\centering
%\fbox{\includegraphics[scale = 0.265]{Fig2_reconstruction.png}}
%\caption{Amplitude of the reconstructed wavefronts at the pinhole plane for the (220) reflection of a \SI{100}{\micro\meter} Si thin crystal at \SI{8.346}{\kilo\electronvolt} the two scan regions of the (a) and (b).}
%\label{fig:reconstructions_pinhole_plane}
%\end{figure}
%Discuss the non-strain data

%\begin{figure}[htbp]
%\centering
%\fbox{\includegraphics[scale = 0.35]{Fig3_propagation.png}}
%\caption{(a) Propagation of the reconstructed wavefronts to the focus of the FZP to focus. (b) Projection of the amplitude of the signal along the transverse direction to diffraction.}
%\label{fig:Propagations_focus}
%\end{figure}
 
In Fig. \ref{fig:reconstruction_all}(a) we show the ptychographic reconstruction of the diffracted wavefront scanned at the pinhole plane.
The range of these two scans was enough to cover the full \SI{78}{\micro\meter} of the diffracted signal, it is possible to observe that the signal in the pinhole plane is elongated in the horizontal, $x$, direction and presents a distribution of maxima, the echoes.
Due to the divergence of the beam, the wave-field at the pinhole position extends a few microns along the vertical, $y$, direction. 
The drop in intensity in the center along the $y$ direction is due to the beam-stop after the FZP.
Because the reconstruction algorithms retrieve the amplitude and phase of the x-ray field produced at the pinhole position.
It is possible to propagate the wave-field back to the focal plane, \SI{5}{\milli\meter} upstream of the pinhole after cropping the reconstructed signal to remove the edges of the reconstruction.
The reconstructed image at the focal plane (back surface of the sample) is shown in Fig. \ref{fig:reconstruction_all}(b).
At this plane, the spatial resolution to resolve the echoes is the highest and a total of 10 echoes are observed, including those at the edges of the Borrmann fan.
These outermost echoes are related to the asymptotes of the dispersion curves of dynamical diffraction \cite{Batter64,Bat68,Authi01} (at the extreme angles of the accepted illumination).
%the focus plane the spatial resolution is higher, we have chosen the depth of focus to match the thickness of the crystal and in that way we can assume that the rear surface in which the echoes are generated lies on the focal plane. 
%A total of 10 echoes are observed, also the ones located at the extremes of the Borrmann fan.
%These two echoes relate to the diffraction from the two surfaces of the crystal. 
The x-ray beam at the focus extends over \SI{73}{\nano\meter} in the $y$ direction, while due to the dynamical diffraction the x-ray footprint in the $x$ direction extends around \SI{78}{\micro\meter}.
%which relates to the finger print of the x-rays in the Borrmann fan at the rear surface of the crystal in Laue diffraction geometry.
Fig. \ref{fig:reconstruction_all}(d black stars) shows the sum along the $y$ axis of the propagated signal to the focus as presented in Fig. \ref{fig:reconstruction_all}(b).

The intensities of the echoes in the diffraction direction are all of the same order of magnitude.
In comparison,in our previous work for the forward direction \cite{ARF21}, we observed that the damping effect is greater, such that the echo maxima variation was approximately one order of magnitude between contiguous echoes.
%over one order of magnitude observed was 1 order of magnitude smaller than the forward beam
This relates to the lack of excitation in one of the branches of the dispersion curve in the forward direction, while in the Laue geometry both diffraction branches are excited.
In Fig. \ref{fig:reconstruction_all} (b) instead of the linear modulation along the x direction as present in Fig. \ref{fig:reconstruction_all} (c) it is possible to observe a tilt in the signal that presents an extra modulation along the y direction.
This may also be visible if we perform reconstructions in each scan separately, as shown in the SM Fig. S6 for a similar scan, which raised doubts about the alignment of the reflection roll with the detector.
Moreover, this artifact appears stronger when stitching the two scans, probably due to an uncertainty in the position between the two stitched scans. 
Attempts to correct a possible misalignment and/or the relative position between the two scans did not change the results.
However, the integrated intensity seems to match well the simulations as presented in the Fig. \ref{fig:reconstruction_all}(c)
%the reconstructed wave-front presented in (a) it is not possible to observe a misalignment of the sample or scanner, in (b) it is possible to observe a tilt in the signal.
%We do not think that this is related to a misalignment of the sample diffraction geometry, as the single scan reconstruction for each of the scans is not presenting this tilt.
%In other scans where the sample was re-aligned the signal looks to be horizontally straight, but this scan did not have enough overlap to perform the stitching reconstruction.
%We present one of them in the Support Material Fig. (XX)

%Table of positions?
\begin{table}[htbp]
\centering
\caption{ Spatial displacement ($\Delta$x), intensity, FWHM and time delay calculated using equation \ref{eq:refShvydko}}
\begin{tabular}{ccccc}
\hline
Maxima & $\Delta$x($\mu$$m$) & I($arb. units$) & FWHM($\mu$$m$) & $\Delta$t($fs$)\\
\hline
$1$  &  0     & 0.72  &  0.93 &   0    \\
$2$  &  1.42  & 0.70  &  1.55 &   1.98 \\
$3$  &  5.24  & 0.50  &  3.13 &   7.33 \\
$4$  & 11.97  & 0.38  &  5.57 &  16.73 \\
$5$  & 28.36  & 0.40  & 20.53 &  39.66 \\
$6$  & 49.12  & 0.44  & 18.49 &  68.69 \\
$7$  & 65.31  & 0.35  &  5.19 &  91.33 \\
$8$  & 72.17  & 0.53  &  3.22 & 100.93 \\
$9$  & 75.95  & 0.64  &  1.62 & 106.21 \\
$10$ & 77.53  & 0.80  &  0.79 & 108.42 \\
\hline
\end{tabular}
  \label{tab:recons-delay}
\end{table}

In Table \ref{tab:recons-delay} we present the numerical values extracted from the data using a fit to the curve shown in Fig. \ref{fig:reconstruction_all}(d).
Each of the 10 visible maxima was fitted using a Gaussian distribution for each of the peaks as presented in Fig. S5 of the SM.
We denote the first echo as the first maxima of the exit beam along the diffraction direction, $k_H$, as represented in Fig. \ref{fig:Geometry}.
This would also be the position of the diffracted beam in the case when only the front surface was diffracting the x-rays.
The spatial displacement, $\Delta$x, of the different maxima is presented in reference to the first echo together with the value of the intensity and full width half maxima (FWHM).
This $\Delta$x is related to a temporal delay generated in the crystal, $\Delta$t, as presented in the work by Lindberg and Shvydko \cite{Shvy12,Lin12}:

\begin{equation}
\Delta x = c \cot(\theta_B) \Delta t
\label{eq:refShvydko}
\end{equation}
where c is the speed of light and $\theta_B$ the diffraction angle.
In the fourth column of table \ref{tab:recons-delay}, we show the value of the expected $\Delta$t using eq. (\ref{eq:refShvydko}).
The zero time delay is associated with the first maximum $(\#1)$, as it is the one that has the shorter trajectory to the detector plane (see Fig. \ref{fig:Geometry}).
This effectively means that, if one used an infinitely short x-ray excitation, by spatially selecting one single echo, or a limited region of the echoes distribution, one could produce a pulse with controlled duration and time of arrival.  
%In practice, this can be obtained using an x-ray pulse smaller than the width of the chosen echo, cf. Fig. \ref{fig:simulations_time}(a).

%\begin{figure}[htbp]
%\centering
%\fbox{\includegraphics[scale = 0.27]{Fig4_simulations.png}}
%\caption{(a) Simulation of the diffracted signal at the rear surface of the crystal in the x-y plane for a Si (220) \SI{100}{\micro\meter} crystal at \SI{8.346}{\kilo\electronvolt} with a beam size of \SI{100}{\nano\meter}. 
%(b) Integrated intensity of the simulated signal along the $x$ direction.}
%\label{fig:Simulations}
%\end{figure}

Fig. \ref{fig:reconstruction_all}(c) shows a simulation of the expected signal in the x-y plane at focus while using an x-ray beam width \SI{500}{\nano\meter} waist.
The simulation was performed using the same code as described in \cite{ARF18,ARF20,ARF21} for a Si \SI{100}{\micro\meter} set to diffract at \SI{8.346}{\kilo\electronvolt} for the symmetric $(220)$ Laue reflection in horizontal geometry.
This simulation output is a 3D matrix containing $x$, $y$ directions and time, $t$.
The simulated x-ray beam has a constant phase and energy bandwidth defined by a simulated Si $(111)$ monochromator.
Fig. \ref{fig:reconstruction_all}(d) presents the sum along the $y$ axis of the signal, which allows the reader to better compare the experimental and simulated intensities.
In Fig. \ref{fig:reconstruction_all}(d), the gray dots represent the intensity profile of the propagated reconstructions, while the black dashed line presents the profile of the simulated data.
In this plot, the two curves have been normalized to the first maximum.
It is possible to observe a high agreement of the position of the diffracted maxima for both signals.
%One can observe that the $\Delta$x of the echoes for the simulated data matches the $\Delta$x of the propagated reconstructed data.
The intensity of the maxima in the center of the reconstructed wavefront data looks to be higher than in the simulations, which could be an effect of the divergence of the incident beam in the sample. 
%Comparing both, simulation and reconstructed data, it could be though that divergence spreads the output radiation toward the center of the Borrmann Fan.
In Fig. \ref{fig:reconstruction_all}(d), it is possible to observe in the propagated reconstructed profile a higher frequency modulation on top of the reconstructed echoes, this could be related to the effect of residual strain on the crystal surface due to the processes of cutting and polishing or imperfections along the thin crystal depth as presented in \cite{ARF21}.
Nevertheless, noise in the reconstruction can not be discarded.

\begin{figure}[htbp]
\centering
\fbox{\includegraphics[scale = 0.7]{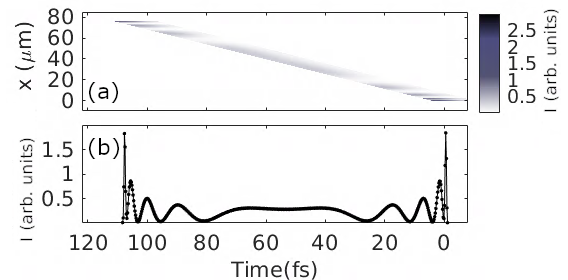}}
\caption{(a) Spatio-temporal simulation of the diffracted signal for a Si crystal of \SI{100}{\micro\meter} set for the (220) Laue geometry reflection at \SI{8.346}{\kilo\electronvolt} with a beam size of \SI{100}{\nano\meter}, using a monochromatic beam with a pulse length of \SI{10}{\femto\second}. 
(b) Projection of the intensity of the simulated signal along the time direction.}
\label{fig:simulations_time}
\end{figure}

In Fig. \ref{fig:simulations_time}(a) we present the spatio-temporal coupling simulation.
This result is the projection of the 3D matrix onto the plane formed by time and the $x$ direction.
One can observe that the positions of the maxima follow the Shvydko-Lindberg equation presented in eq. (\ref{eq:refShvydko}).
From these results, it can be extracted that the total temporal difference between the two exterior echoes 1 and 10 is less than \SI{110}{\femto\second}.
The projection to the temporal axis is presented in Fig. \ref{fig:simulations_time}(b).
The temporal simulation shows that the two diffraction echoes resulting from the two maxima at the two extremes of the fan are delayed between each other, even if the length of the path through which the photons travel inside the crystal is the same.
The photons coming from echo 10 must travel an extra distance in the free space after leaving the crystal surface with respect to the photons in echo 1, as presented in Fig.\ref{fig:Geometry}.
As we acquired a ptychographic scan that takes a fraction of a second per scan point, the time delay of the echoes between each other can only be inferred by simulations.
It would be very interesting to perform a temporal study to corroborate these temporal simulations.
Perhaps such an experiment could be possible at an XFEL with a fast enough timing tool and probably using a short enough x-ray pulse (in the attoseconds range) to properly distinguish between echoes; in combination with a single shot reconstruction technique as described by Lee and co-workers \cite{Lee23}.
%This could be used to measure the time in which the diffraction process is happening for each thickness of the crystal.

%Showing delays of few $fs$ between them as presented in table \ref{tab:recons-sim-delay}.

\begin{figure}[htbp]
\centering
\fbox{\includegraphics[scale = 0.6]{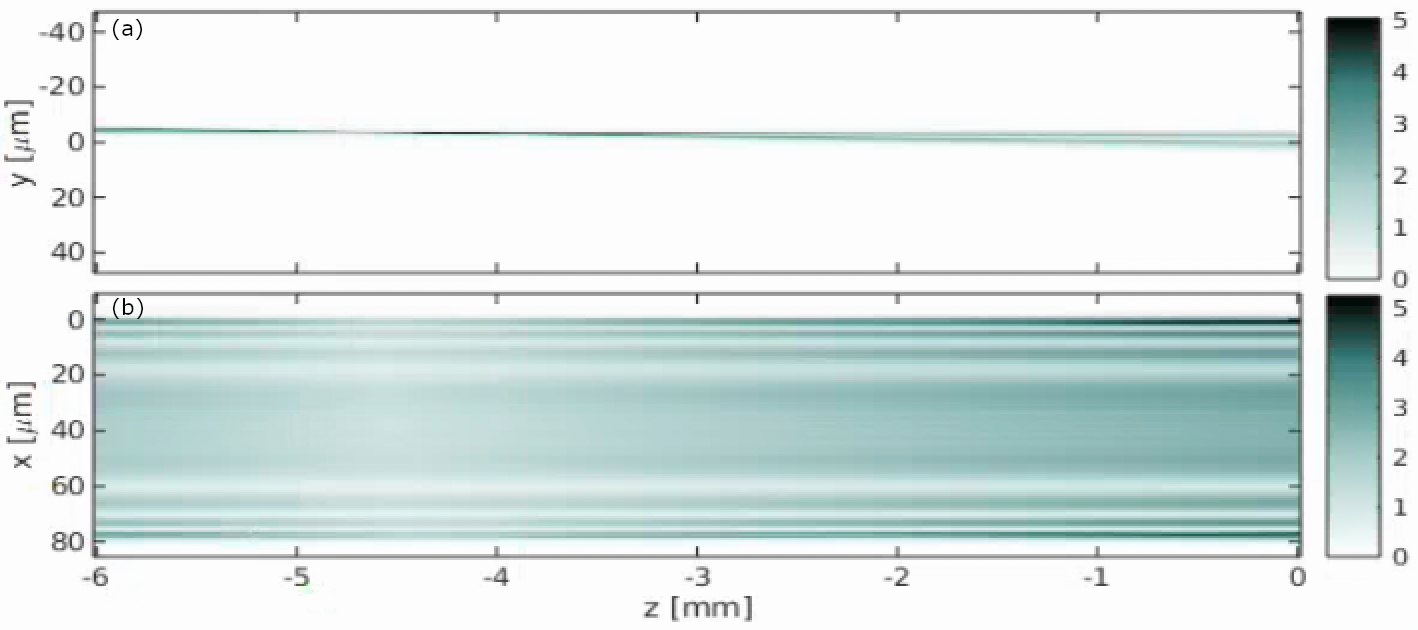}}
\caption{Propagation of the reconstructed x-ray diffracted beam along the $z$ direction (a) for the vertical plane and (b) the horizontal plane where diffraction takes place. 
}
\label{fig:Propagation_z}
\end{figure}

In Fig. \ref{fig:Propagation_z}, we present the propagation along the $z$ direction for both vertical ($y$-$z$) and (b) horizontal ($x$-$z$) plane.
While for the vertical plane, Fig. \ref{fig:Propagation_z} (a), we observed that the beam propagation is dominated by the divergence of the focused beam, for the horizontal plane, Fig. \ref{fig:Propagation_z} (b), it is possible to see how the propagation of the echoes from the focus to the pinhole plane is almost unaffected by the divergence of the beam.
This confirms that the echoes are parallel diffracted beams, as is known from theory and simulations.
As highlighted in Ref \cite{ARF18} and Ref\cite{ARF21}, the Si crystal behaves like a monochromator, selecting a limited portion of the divergent beam.
%propagates according to what would be expected from a normal divergent focus beam.

%And interesting observation that is open to discussion is the following: if the distance that the photons for the beam 1 and 10 inside the crystal is the same, why should be that the beam 10 delay with respect to the beam 1 around \SI{108}{\femto\second}. 
%Could not be possible that these two beams have the same zero delay and that echoes 5 and 6, the ones in the middle of the range have the longer delay. 
%From the studies performed in the forward direction, one could support the almost \SI{108}{\femto\second} delay between echoes 1 and 10, but they only real answer is to perform a timing experiment at a facility with femtosecond temporal resolution as an XFEL. 

%%%%%%%%%% ideas for discussion
%Talk about the idea of using the pendolosium effect to do time studies, change wavelength will change the position of the echoes fringes.

% the echoes to narrow the temporal beam distribution using a pinhole or slit.
%%%%%%%%%%%%%

\section{Conclusions}
In this work, we have taken our research a step further, by imaging the Borrmann fan echoes produced by a crystal in Laue geometry in the diffraction direction with a resolution of about \SI{100}{\nano\meter}.
%We have observed that for a Si crystal with thickness \SI{100}{\micro\meter} at an energy of \SI{8.346}{\kilo\electronvolt}, set to diffract in the (220) Laue geometry reflection, t
A total of 10 x-ray beams have been observed for the chosen sample and experimental conditions.
The echoes present in the Borrmann fan extend for around \SI{78}{\micro\meter}, these echoes are observed to have photon intensities of similar magnitude, as expected. 
The spatio-temporal coupling of the echoes would translate to a maximum time delay between the first and the last echoes of around \SI{108}{\femto\second}, and a time delay shorter than \SI{2}{\femto\second} for the closest echoes.
Our measurement at a synchrotron source are static and the echoes can be resolved due to the lateral displacement in the Borrmann fan. To resolve the temporal delay between the different beams, both timing experiments based on pump-probe schemes or an interference experiment based on the different path between the echoes could be performed at a facility with femtosecond time resolution.

These diffracted beams are present for both diffraction geometries, Laue and Bragg, as presented in \cite{ARF18}.
When thin perfect crystals are used for x-ray optics, the echoes can generate disturbances in the incoming wavefront.
The pendell\"{o}sung effect was observed in thin InSb crystal structures \cite{verezhak2020}, which would suggest that there is enough thickness to observe the echoes for such a material.
For this to happen, the thickness of the sample must be similar to the extinction length for that energy and reflection.
As presented in our previous work, strain can affect the structures of the echoes \cite{ARF21}, what could create complex wavefronts in the case of studying thin crystals from elements with high Z such as Ni, Au or InSb.

The dynamical diffraction echoes could be used at XFELs facilities to temporally split beams with delays of a few femtoseconds between each other.
These beam splitters could be used to study ultrafast processes using techniques such as XPCS  \cite{Anders03,Jo2021,Reiser22}.
This delay relates to processes in the THz frequency, suggesting that thin crystals could be used as beam splitters to study ultrafast processes in the THz frequencies.
By selecting crystal thickness, reflection and x-ray energy, the delay between the echoes could be arranged to study different temporal lengths.

In addition, because of the dependence of the signal on the distortions of the crystal lattice in depth, the echoes could be used as a streaking method to resolve ultrafast temporal distortions in single crystals. 
We note that the echoes propagate in depth with the speed of light, while the distortions are traveling at a much slower speed closer to the speed of sound.
In a recent work, the echoes have been used to study the distortion process in Si after femtosecond laser excitation \cite{ARF25}. 
In that work, the authors had to locate the sample out of focus to increase the resolution at the detector plane.
The effect of the divergence beam affected the signal, producing that the echoes next to the signal edges overlapped and in that way the temporal resolution due to the echoes for short time delays was lost. 
Here, we have presented that tele-ptychography could resolve these fringes and allow the sample to be in focus.
In our next steps in the evolution of tele-ptychography, we want it to become a single pulse resolved technique.
Several groups are working in this direction, in particular the coherent speckle technique presented by Lee and co-workers looks as a good approach \cite{Lee23}.
This development could be not only used in temporal studies of Ge, Si or diamond as presented in \cite{ARF25}, but extended to study other thin crystals with shorter extinction lengths such as InSb or Au, in which the time delays between echoes would be shorter than the femtosecond as presented in \cite{ARF23}.

%Introduce how the use of this echoes can be used to have temporal resolutions interestign also in the future sources. (future optics for  bellow 10 fs)
%ultrafast optics

%THz studies
%maybe talk about pendolosium and change of wavelength

% Prsent the results an the relevance for the future of the x-ray sources

%\section{References}

%Note that \emph{Optics Letters} and \emph{Optica} short articles use an abbreviated reference style. Citations to journal articles should omit the article title and final page number; this abbreviated reference style is produced automatically when the \emph{Optics Letters} journal option is selected in the template, if you are using a .bib file for your references.

%However, full references (to aid the editor and reviewers) must be included as well on a fifth informational page that will not count against page length; again this will be produced automatically if you are using a .bib file.

%\bigskip
%\noindent Add citations manually or use BibTeX. See \cite{Zhang:14,OSA,FORSTER2007,testthesis,manga_rao_single_2007}.

%\appendix % if required
%\section{Appendix title}

%Text text text text text text text text text text text text text text
%text text text text text text text.

\begin{acknowledgements}
%M.V. acknowledges funding by European Union's Horizon 2020 research and innovation program under the Marie Sk\l{}odowska-Curie grant agreement No 701647, and the SNSF grant No 200021L\_169753.
We acknowledge the European Synchrotron Radiation Facility (ESRF) for provision of synchrotron radiation facilities  under proposal number MI-1398 and we would like to thank the ID01 team for assistance and support in using beamline ID01.
This research was supported in part through the Maxwell computational resources operated at Deutsches Elektronen-Synchrotron DESY, Hamburg, Germany.
One of us would like to thank K. Finkelstein for many hours of discussion about dynamical diffraction and showing the first steps in the world of pump-probe in single crystals.
We thank M. Verezhak, Zdenek Matej, Manuel Guizar-Sicairos, Bill Pedrini, Virginie Chamard and Kenneth Finkelstein for useful discussions.

 A.R.F and A.D. conceptualized the work; A.R.F., A.D. and D.C. planned the experiment;  A.R.F., D.C, D.K. and A.D. performed the experiment; A.R.F. analysed the data with contributions of A.D. and D.K.; A.R.F. performed the simulations; A.D. and A.R.F. wrote the manuscript with the contribution of all authors.
\end{acknowledgements}

%\DataAvailability{Please state how the data supporting the results reported in your article can be accessed, e.g. within the article, as published supporting material, in repositories, upon request...
%}

%\bibliography{ESRF_Si_echoes} % basename of .bib file

\end{document}